# Discovery of Superoxide Reductase: an Historical Perspective

Vincent Nivière . Marc Fontecave

Laboratoire de Chimie et Biochimie des Centres Redox Biologiques, DRDC-CB, UMR CEA/CNRS/Université Joseph Fourier n° 5047, CEA Grenoble, 17 Avenue des Martyrs, 38054 Grenoble Cedex 9, France

V. Nivière (✉) . M. Fontecave (✉)

VN. E.mail: vniviere@cea.fr; Tel: 33-4-38789109; Fax: 33-4-38789124

MF. E.mail: mfontecave@cea.fr; Tel: 33-4-38789103; Fax: 33-4-38789124

Abbreviations

SOD, superoxide dismutase; SOR superoxide reductase; Dfx, desulfoferrodoxin

Abstract

For more than thirty years, the only enzymatic system known to catalyze the elimination of superoxide was superoxide dismutase, SOD. SOD has been found in almost all organisms living in the presence of oxygen, including some anaerobic bacteria, supporting the notion that superoxide is a key and general component of oxidative stress.

Recently, a new concept in the field of the mechanisms of cellular defense against superoxide has emerged. It was discovered that elimination of superoxide in some anaerobic and microaerophilic bacteria could occur by reduction, a reaction catalyzed by a small metalloenzyme thus named superoxide reductase SOR.

Having played a major role in this discovery, we describe here how the concept of superoxide reduction emerged and how it was experimentally substantiated independently in our laboratory.

*Introduction*

The general question of the toxicity of molecular oxygen present at the surface of the earth continues to be a major topic in modern biology. It is now well established that all living organisms, both aerobes and anaerobes, have developed mechanisms to protect themselves specifically from this toxicity [1, 2]. In some cases oxygen can be deleterious by itself, but the most reactive species are those derived from reduction of oxygen: the superoxide radical, hydrogen peroxide and the hydroxyl radical [1, 2]. During the last thirty years biochemists and biologists working in this field have thus focused their attention basically to the three following problems. The first concerns the identification of the toxic species, the biological targets and the molecular mechanisms by which an oxidative stress is expressed within an organism. A second question concerns the role played by such an oxidative stress in aging and in a number of diseases (neurological disorders, some types of cancer, inflammation,…) and the design of suitable therapeutical strategies. The third problem concerns the nature and mechanisms of the complex antioxidant machinery upon which living organisms rely for controlling the balance between the generation and the scavenging of the reactive oxygen species, continuously (aerobes) or transiently (anaerobes). All three topics have been the subject of review articles in the recent past [2-6] and it is not our intention to add one to the list. In this brief paper, we want to tell the story of an important discovery which, a few years ago, changed our current views on the biological mechanisms of superoxide detoxification.

Before the end of the 90's, it was established that the unique biological mechanism for scavenging superoxide radicals, $O_2^{\bullet-}$, was by dismutation to molecular oxygen and hydrogen peroxide, a reaction (eq. 1) catalyzed by a metalloenzyme thus named superoxide dismutase (SOD):

$$O_2^{\bullet-} + O_2^{\bullet-} + 2H^+ \rightarrow H_2O_2 + O_2 \quad (1)$$

The first SOD enzyme was discovered in 1969 by McCord and Fridovich [7]. It has been then established that all organisms living in the presence of oxygen contain at least one of the four possible types of SODs, which differ by the metal center present in the active site, namely manganese, iron, nickel or copper-zinc [8, 9]. The fact that in aerobic life a defense enzyme is specifically required to destroy $O_2^{\bullet-}$ demonstrates that this species is sufficiently toxic by itself or indirectly through its conversion to more harmful compounds. In a great variety of living organisms, inactivation of *sod* genes has been shown to perturb the viability of the cells dramatically [10, 11]. The use of the SOD protein itself or synthetic compounds (SOD mimics) for protecting tissues from oxidative injury due to superoxide has thus received increased attention [12].

The discovery of the presence of SOD enzymes also in strictly anaerobic bacteria further supported the notion that $O_2^{\bullet-}$ is a key and general component of oxidative stress [13, 14]. These organisms may be accidently and transiently exposed to oxygen and the availability of antioxidant mechanisms may be an advantage [15]. In fact, it has been recently discovered that some microaerophilic and anaerobic bacteria use not only SOD but also a new class of superoxide-scavenging enzymes, named superoxide reductases (SOR) [16-19], which catalyze the following reaction (eq. 2):

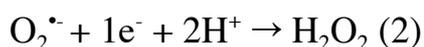

$$O_2^{\bullet-} + 1e^- + 2H^+ \rightarrow H_2O_2 \quad (2)$$

These systems have been purified to homogeneity and extensively characterized by a growing number of research groups which provided a wealth of important information regarding their spectroscopic, structural and reactivity properties. In 2002, a series of commentaries was published in the Journal of Biological Inorganic Chemistry in order to take stock of the current knowledge on these enzymes [20-24]. As a supplement to the information provided by these articles, we would like to describe how the concept of superoxide reduction emerged and how it was independently experimentally substantiated in our laboratory [17], essentially simultaneously with related work in Athens, Georgia [16].

*Superoxide scavenging in sulfate-reducing bacteria*

The first observation was provided by Danièle Touati in Paris. In a paper published in 1996 [25], she reported her attempts to isolate a *sod* gene from *Desulfoarculus baarsii*, in order to understand the origin of the considerable aerotolerance of some anaerobic sulfate-reducing bacteria. She showed that functional complementation of the *sod⁻ recA⁻* deficient *Escherichia coli* mutant strain with *D. baarsii* genes unexpectedly led to the isolation of the *rbo* gene. This gene had been fortuitously sequenced in *Desulfovibrio vulgaris* Hildenborough, another sulfate-reducing bacteria, in 1989 by Gerrit Voordouw [26]. *Rbo* was obviously different from any other known *sod* genes. The name of this gene (*rbo* for rubredoxin oxidoreductase) was tentatively given on the basis of the presence, downstream of the *rbo* gene within the same operon, of a *rub* gene encoding a rubredoxin and the coordinated expression of the two genes [26]. However, it should be noted that the association of *rbo* and *rub* genes is quite unique to *D. vulgaris* and *D. baarsii* and is not a general feature of microbial genomes. Thus it is likely that this gene was originally misnamed, even though rubredoxin was recently shown to be able to transfer electrons to Rbo [27].

In the mid-90s, it also became obvious that the Rbo protein was identical to desulfoferrodoxin (Dfx), a protein isolated from *Desulfovibrio desulfuricans* and *D. vulgaris* by Isabel and José Moura and Jean Le Gall in Lisbon. In two excellent papers published in the *Journal of Biological Chemistry* in 1990 and 1994 [28, 29], these groups in collaboration with Michael Johnson (Athens, USA) and Boi Hanh Huynh (Atlanta) reported a thorough characterization of Dfx by a variety of spectroscopic methods including EPR, Raman resonance, Mössbauer spectroscopy. In 1997, the determination of a three-dimensional structure of Dfx from *D. desulfuricans* by M.A. Carrondo (Oeiras, Portugal) [30] confirmed all the spectroscopic predictions, in particular the presence of two separate mononuclear non-heme iron centers per polypeptide chain. Center I is a ferric site with a distorted tetrahedral sulfur coordination and center II is a ferrous site with a unique square pyramidal structure involving one cysteine and four histidine residues as ligands (Figure 1). The name of the protein, desulfoferrodoxin, again was given not on the basis of its function, since no biological activity could be assigned to Dfx at that time, but on the basis of the presence of a ferric site (center I) similar to that of desulforedoxin [31], a rubredoxin-like protein from *Desulfovibrio* species, together with a ferrous site (center II).

*Rbo-Dfx is not a SOD enzyme*

In 1997, we started a close collaboration with Danièle Touati with the aim of identifying the activity of Rbo-Dfx. We soon came up with two obvious working hypotheses. Rbo-Dfx could possibly be a novel SOD and thus catalyze superoxide dismutation. This was not unlikely since superoxide dismutase activity is not restricted to a single class of metalloprotein, as indicated above. Using the *D. baarsii* Rbo-Dfx expressing plasmid provided by D. Touati, with the help of a young student, Nelly Minguez, we rapidly got pure preparations of the protein by the beginning of 1997 and thus were able to evaluate its SOD activity. Using different SOD assays (inhibition of superoxide-dependent pyrogallol autoxidation or inhibition of nitro blue tetrazolium reduction, both monitored spectrophotometrically), we soon found that Rbo-Dfx had just a very weak SOD activity as compared to real SOD enzymes (the specific activity of Rbo-Dfx corresponded to about 0.5 % of that of SODs). We thus ruled out the hypothesis that Rbo-Dfx was a novel class of SOD. We then were rather surprised in 1999 by a paper from M. Teixeira [32] concluding, in marked contrast with our analysis, that the Rbo-Dfx protein from *D. desulfuricans* was a novel SOD. This conclusion seemed also inconsistent with their own measurements of SOD activity (25-70 units/mg corresponding to about 1 % of the activity of real SOD enzymes). The same year, the same group reported an activity of 1200 units/mg for a protein, named neelaredoxin [33] because of its blue color, which had been isolated from *Desulfovibrio gigas* in 1994 [34]. This protein contains a unique mononuclear non-heme iron center identical to center II of Rbo-Dfx and no center I.

However, the large SOD activity value that was reported may arise from an incorrect interpretation of the results obtained with the cytochrome c assay. This standard SOD assay is based on the inhibition of cytochrome c reduction by superoxide radicals, enzymatically generated with the xanthine/xanthine oxidase system. As demonstrated by M. Adams at the end of 1999 [16] and by us at the early beginning of 2000 [17], the activity determined with this assay, in the case of Rbo-Dfx, does not reflect a true SOD activity. Later on in 2000, M. Teixeira, using pulse radiolysis techniques, confirmed that neelaredoxin from *Archaeoglobus fulgidus* displayed also only a very weak SOD activity [35].

It thus seems clear that Rbo-Dfx and neelaredoxins having an iron center II in common are not novel SODs. However there is no doubt that, as expected for a metal-dependent system, these proteins display a small superoxide dismutase activity of about 20-100 units/mg, much below that of canonical SODs (4000-5000 units/mg). This observation raises the question whether this small activity is sufficient by itself to provide protection against superoxide *in vivo* and, for example, to restore growth of *sod*- *E. coli* mutant strains. In fact, a convincing work by J. Imlay [36] demonstrated that the level of SOD activity in *E. coli* cells is just enough for protection against oxidative stress. *E. coli* cannot tolerate steady-state levels of superoxide significantly in excess of 0.1 nM and as a consequence it requires substantial concentrations of SOD (50 $\mu$M). With such a low SOD activity and the level of expression in *E. coli* (not more than 5% of total soluble proteins) Rbo-Dfx is unlikely to achieve protection against superoxide by dismutation. We thus soon believed that the SOD activity of these proteins was not physiologically relevant, as discussed by us in [17].

*Evidence for Rbo-Dfx as a superoxide reductase*

In 1997, our attention was focused on the intriguing results we obtained with the cytochrome c assay. It took time to realize that in solution producing superoxide, cytochrome c was oxidized (rather than reduced), as shown by the decrease of the intensity of the 550 nm absorption band characteristic of reduced cytochrome c, and that this oxidation was strictly dependent on the presence of Rbo-Dfx. Since the addition of large amounts of SOD had only weak inhibitory effects on the oxidation of cytochrome c, the direct involvement of superoxide in the reaction was not obvious. Later on, we demonstrated that Rbo-Dfx very efficiently competed with SOD for direct reaction with superoxide, in agreement with the observation that cytochrome c oxidation was stoichiometric with the production of superoxide by the xanthine/xanthine oxidase system. We thus by the end of 1997 came to the conclusion that, in these preliminary experiments, Rbo-Dfx behaved as a catalyst of the reduction of superoxide by reduced cytochrome c (Figure 2). All these results were described latter in our first paper [17] and led us to designate Rbo-Dfx as a superoxide reductase.

In a paper published in October 1997 [37], I. Fridovich showed that expression of Rbo-Dfx in the $sod^-$ *E. coli* mutant provided by D. Touati drastically decreased the amount of detectable superoxide (using the luminescent reaction with lucigenin). In the same paper, however with no experimental evidence, the suggestion was made that Rbo-Dfx could protect cells against superoxide radicals by reduction provided that the active reduced form of Rbo-Dfx could be regenerated by cellular reductants such as NAD(P)H or glutathion. In an earlier study, I. Fridovich had shown that indeed some compounds, such as synthetic Mn-porphyrins, could scavange superoxide by reduction (rate constant: $4 \times 10^9$ $M^{-1}s^{-1}$) rather than by dismutation [38].

The name of superoxide reductase, abbreviated SOR, appeared for the first time in the literature in October 1999 in a *Science* paper [16], by M. Adams (University of Georgia, Athens, USA), submitted 18 June 1999. It described preliminary experiments showing the cytochrome c-superoxide oxidoreductase activity (Figure 2) of a neelaredoxin protein from *Pyrococcus furiosus* containing a single iron center identical to the center II of Rbo-Dfx. Our first paper, submitted 19 May 1999, reporting our discovery of the superoxide reductase activity came out in the first January issue of *Journal of Biological Chemistry* in 2000 [17]. From the experimental work of Murielle Lombard, a PhD student and one of us (Vincent Nivière) during the years 1998-1999, we reported in this paper the first direct experimental evidence that!: (i) Rbo-Dfx is a specific and efficient one-electron reductant of superoxide, the reaction taking place with a rate constant of $6\text{-}7\times10^8$ $M^{-1}s^{-1}$; (ii) the ferrous iron of center II is the electron donor and is converted to the corresponding ferric form during the reaction; (iii) microorganisms such as *E. coli* contain reducing activities allowing reduction of the iron center II required for multiple turnovers. Whether there is a specific Rbo-Dfx reductase to maintain the protein reduced and active is still a matter of discussion. Considering the high redox potential of center II [19, 29, 34, 35] we believe that its intracellular reduction is achieved by a variety of different reducing agents, ascorbate and cellular reductases either NAD(P)H-dependent or –independent [17, 27, 29].

*Why such a fast $O_2^{\bullet-}$ reduction reaction ?*

The most remarkable property of SOR is certainly its great efficiency in reacting with superoxide, with a velocity close to the limit of diffusion of a molecule in solution [39-43]. We and others pointed out that the active site is exquisitely designed to provide strong electrostatic attractive power with regard to superoxide [40, 43, 44]. Other still unidentified structural factors are probably implicated in the enzyme-superoxide interaction. Furthermore, a very fast electron transfer reaction is allowed because of the intermediate formation of an iron-superoxide complex. As an evidence for an inner-sphere mechanism, intermediate iron-peroxo species have been proposed from DFT calculations [45] and observed by resonance Raman spectroscopy in the active site of SOR [46].

The requirement for such a rapid reaction is intimately associated with the specific chemical properties of superoxide. The fact that superoxide (i) dismutates spontaneously with a rate constant of about $5 \times 10^5$ $M^{-1}$ $s^{-1}$ in water at pH 7.0 [2] (and probably faster in the presence of trace metals) and (ii) reacts with [4Fe-4S] centers of dehydratases with rate constants in the $10^7$-$10^8$ $M^{-1}$ $s^{-1}$ range [47, 48], implies that any efficient superoxide-scavenging system should interact with superoxide with a rate constant of at least $10^8$ $M^{-1}$ $s^{-1}$. This is indeed achieved by both SOD and SOR enzymes. Furthermore, as a consequence of the very low steady-state concentration of superoxide in the cell ($10^{-10}$ to $10^{-9}$ M), the overall enzyme reaction is not rate-limited by the reduction of SOR by cellular reductases, but instead by the oxidation of SOR by superoxide, as nicely demonstrated by D. Kurtz and coworkers [49].

Then, this unique $O_2^{\bullet-}$ scavenging reactivity is one important feature that future SOR mimics [50] should display if one wants develop them as alternatives to SOD mimics for therapeutical treatment against diseases in which superoxide plays an important role.

Another important question which remains to be unraveled is why some organisms use SOR instead of SOD to detoxify superoxide. It has been proposed that the main advantage of SOR as compared to SOD resides in the fact that SOR does not produce $O_2$, which is expected to be toxic by itself for air-sensitive microorganisms [16]. Although this hypothesis provides a simple clue to explain the presence of SOR in some specific cells, we believe that there are more subtle reasons for that [17, 18]. In fact, anaerobic bacteria, and in particular sulfate-reducing bacteria, are fully crowded with strongly auto-oxidizable soluble redox proteins, such as redox carriers (ferredoxin, cytochromes, rubredoxin, desulforedoxin, flavodoxin) or enzymes, like hydrogenases [51]. Because these proteins are prone to release their electrons, a strong superoxide stress can be generated upon exposure to $O_2$, partly explaining why anaerobes are so sensitive to air [51]. Such a process is probably less important in aerobic cells, which have evolved by integrating the electron transport proteins into the membrane in order to minimize such auto-oxidation reactions [2].

Because of the low specificity of SOR active site for reductants, we believe that these soluble redox proteins can also provide electrons to SOR efficiently. Then, as illustrated in Figure 3, by shuttling the electrons from these auto-oxidizable redox proteins to superoxide, SOR could, in a single reaction, eliminate both superoxide and the source of its production. In addition, such a reaction may allow the anaerobic bacteria to shut off transitory $O_2^{\bullet-}$ production, with no need for sophisticated regulatory systems, such as found in facultative anaerobes.

*Epilogue*!

To our opinion two papers have been at the origin of the concept of superoxide reduction by the Rbo-Dfx and neelaredoxin proteins!: one by M. Adams [16] and one by ourselves [17]. Our paper provided the key information regarding the reaction between center II and superoxide. From then, an increasing number of studies from different laboratories have been carried out aiming at better characterizing the spectroscopic and structural properties of SORs, the reaction with superoxide and the mechanism of the reaction. They nicely confirmed the initial hypothesis and observations. Most of the important results have been presented in the series of commentaries published in 2002 in the Journal of Biological Inorganic Chemistry [20-24].

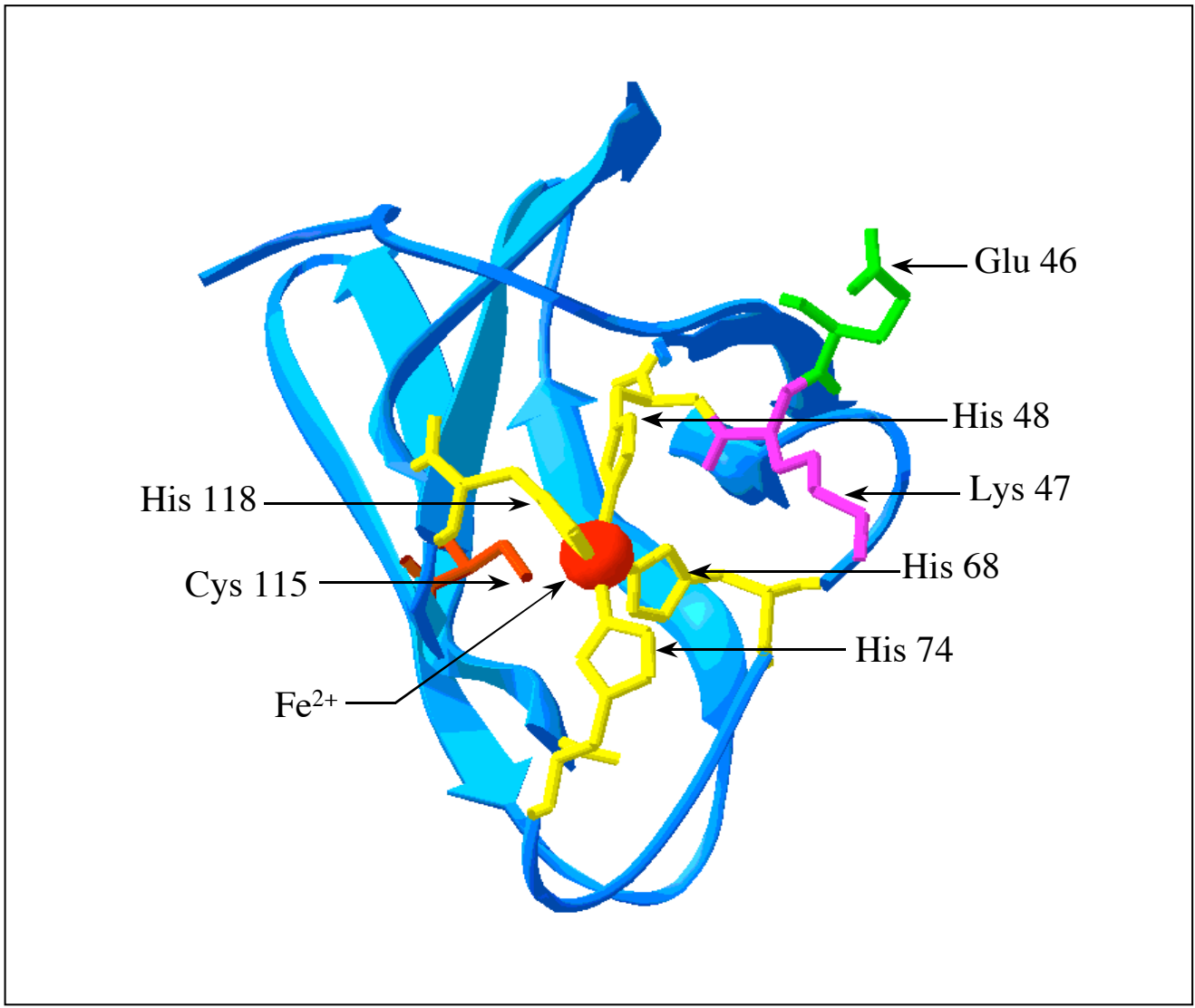

Figure 1. Structure of the ferrous iron center II of *Desulfovibrio desulfuricans* desulfoferrodoxin. Modified from [30].

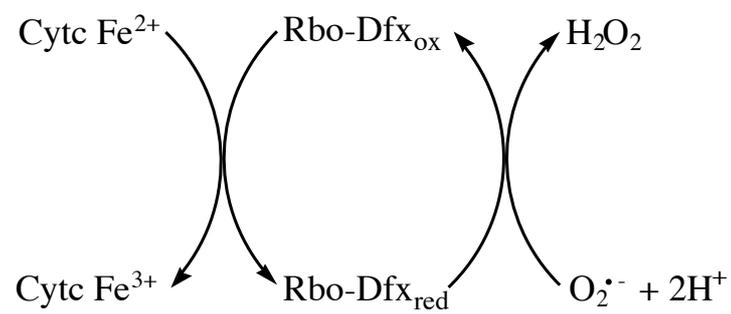

Figure 2. Rbo-Dfx as a catalyst of the reduction of superoxide by reduced cytochrome c.

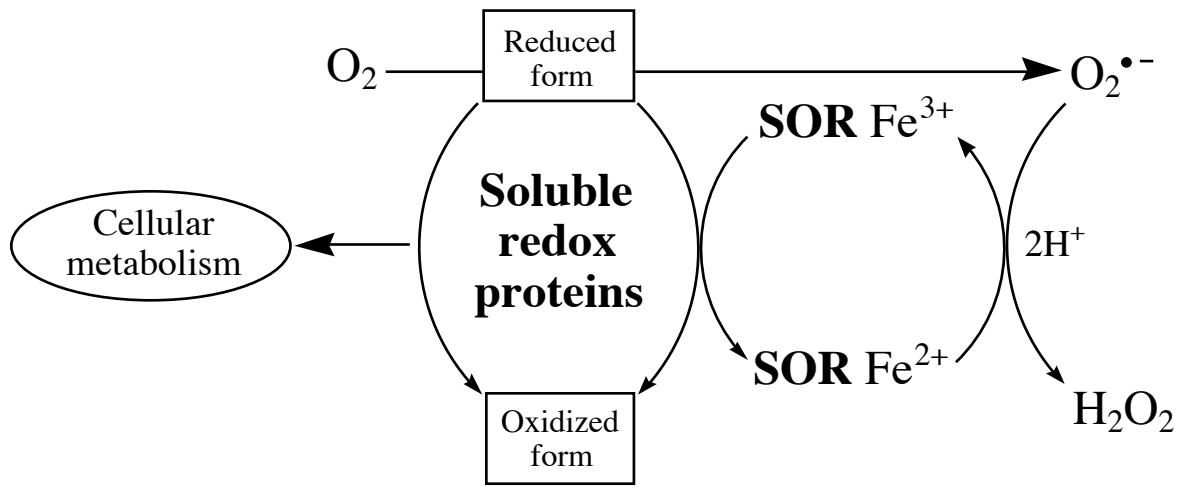

Figure 3. Hypothesis for the detoxification activity of SOR. In the presence of $O_2^{•-}$, formed from the auto-oxidizable redox proteins in the presence of $O_2$, SOR eliminates both $O_2^{•-}$ and its source of production. In the absence of $O_2/O_2^{•-}$, the electrons are shuttled towards the cellular metabolism.